%% file: main.tex
\documentclass[conference,compsoc]{IEEEtran}
\IEEEoverridecommandlockouts
\AtBeginDocument{%
  \providecommand\BibTeX{{%
    \normalfont B\kern-0.5em{\scshape i\kern-0.25em b}\kern-0.8em\TeX}}}

\usepackage{color,soul}
\usepackage{algorithm,amssymb}
\usepackage{algpseudocode}
\usepackage{amsmath,amsthm}

\usepackage{subcaption}
\input{formatting}
\usepackage{graphicx}
\usepackage{gensymb}
\usepackage{booktabs}

\let\oldthefootnote\thefootnote

\renewcommand{\thefootnote}{$\dagger$} 

\makeatletter
\AtBeginDocument{\let\hl\@firstofone}
\makeatother
\begin{document}

\title{Inference Attacks for X-Vector Speaker Anonymization}
\author{\IEEEauthorblockN{Luke A. Bauer}
\IEEEauthorblockA{\textit{University of Florida} \\
lukedrebauer@ufl.edu}
\and
\IEEEauthorblockN{Wenxuan Bao}
\IEEEauthorblockA{\textit{University of Florida} \\
wenxuanbao@ufl.edu}
\and
\IEEEauthorblockN{Malvika Jadhav}
\IEEEauthorblockA{\textit{University of Florida} \\
jadhav.m@ufl.edu}
\and
\IEEEauthorblockN{Vincent Bindschaedler}
\IEEEauthorblockA{\textit{University of Florida} \\
vbindschaedler@ufl.edu}

}

%

\maketitle


\begin{abstract}
  \input{abstract}
\end{abstract}

\begin{IEEEkeywords}
Speaker Anonymization, Inference Attacks, Machine Learning
\end{IEEEkeywords}


\input{intro}

\input{background}

\input{problem}
\input{methodology}

\input{results}

\input{limitations}

\input{conclusions}

\section*{Acknowledgments}
This work was supported in part by the National Science Foundation under CNS-1933208. Any opinions, findings, conclusions, or recommendations expressed in this material are those of the authors and do not necessarily reflect the views of the National Science Foundation.

\bibliographystyle{IEEEtran}
\bibliography{references}

\appendices


\end{document}

%% file: formatting.tex
\PassOptionsToPackage{breaklinks = true,unicode = true,urlcolor = blue,colorlinks = true,citecolor = red,linkcolor = red}{hyperref}
\usepackage{hyperref}

\usepackage{bm}

\usepackage[nameinlink,capitalize]{cleveref} 

\usepackage[font=small]{caption}

\newcommand{\paragraphb}[1]{\smallskip\noindent{\bf #1.} }

\usepackage{multirow}
\usepackage{array}
\newcolumntype{P}[1]{>{\centering\arraybackslash}p{#1}} 

\usepackage{enumitem}

\usepackage[nameinlink,capitalize]{cleveref} 

\newcounter{mynote}[section]

\newcommand{\thenote}{\thesection.\arabic{mynote}}

\newcommand{\wbnote}[1]{\ifx\outforreview\undefined\refstepcounter{mynote}{\it\textcolor{blue}{(WB~\thenote: { #1})}}\fi}

\newcommand{\vbnote}[1]{\ifx\outforreview\undefined\refstepcounter{mynote}{\it\textcolor{magenta}{(VB~\thenote: { #1})}}\fi}

\newcommand{\lbnote}[1]{\ifx\outforreview\undefined\refstepcounter{mynote}{\it\textcolor{red}{(LB~\thenote: { #1})}}\fi}

\newcommand{\fixme}[1]{\ifx\outforreview\undefined\textbf{\textcolor{red}{[FIXME: #1]}}\fi}

\newcommand{\pr}[1]{\ensuremath{\mathrm{Pr}(#1)}}

\newcommand{\ext}{\ensuremath{{\rm Ext}}}
\newcommand{\synth}{\ensuremath{{\rm Synth}}}
\newcommand{\transf}{\ensuremath{{\rm Transf}}}


%% file: abstract.tex
We revisit the privacy-utility tradeoff of x-vector speaker anonymization. Existing approaches quantify privacy through training complex speaker verification or identification models that are later used as attacks. Instead, we propose a novel inference attack for de-anonymization. Our attack is simple and ML-free yet we show experimentally that it outperforms existing approaches. 

%% file: intro.tex
%
\section{Introduction}
\footnotetext{This paper has been accepted for publication at the 8th IEEE Deep Learning Security and Privacy (DLSP) Workshop 2025.}

\renewcommand{\thefootnote}{\oldthefootnote} 
\setcounter{footnote}{0}  %
The past decade has spurred major advances in user-facing speech processing technologies. This include Automatic Speech Recognition (ASR)~\cite{yu2016automatic}, text-to-speech (TTS) synthesis~\cite{ren2019fastspeech}, and speaker identification or verification~\cite{togneri2011overview}. Another technology that has received recent attention, especially in academic work is speaker anonymization~\cite{fang2019speaker,shamsabadi2022differentially,tomashenko2020voiceprivacy,tomashenko2024voiceprivacy}. The goal of speaker anonymization is to transform speech samples to obscure the identity of the speakers while preserving the content.

A straightforward realization of a speaker anonymization system is to use an ASR model to transcribe the input speech and then synthesize a new audio sample matching the transcription but with a different voice. This approach perfectly protects the privacy of speakers and the resulting audio samples are still useful for some downstream tasks due to preserving the transcripts. However, it sacrifices naturalness in the synthesized voice and discards all of the prosodic features (pitch, emotionality, tone, etc.) present in the original speech samples.

To ensure that we can preserve naturalness and prosody in speech while anonymizing speakers, a number of approaches have been proposed~\cite{fang2019speaker,shamsabadi2022differentially,tomashenko2020voiceprivacy,panariello2024voiceprivacy,tomashenko2024voiceprivacy}. Notably, initiatives such as the VoicePrivacy Challenges~\cite{tomashenko2020voiceprivacy,tomashenko2022voiceprivacy,panariello2024voiceprivacy,tomashenko2024voiceprivacy} have been encouraging development and evaluation of speaker anonymization techniques. One particularly prominent approach is the use of x-vectors~\cite{snyder2018x}, which are embeddings that capture speaker-specific characteristics extracted from speech. Informally, x-vector anonymization schemes transform this x-vector into a {\em pseudo x-vector} using publicly available data (i.e., a public pool of other speakers' x-vectors). This pseudo x-vector is then combined with the transcript and pitch information (extracted from the input speech sample) to synthesize a new audio sample. When the pseudo-x-vector is chosen to be significantly different from the original speaker's x-vector, the synthesized speech sounds natural, as if it was uttered by another person.

However, despite substantial research, the privacy-utility tradeoff of x-vector speaker anonymization is not fully characterized. This is partially due to the fact that previous work has primarily relied on empirical privacy evaluation using speaker identification or speaker verification systems. Said differently, existing approaches train deep learning models to play the role of the attacker. 

Training machine learning models as attacks has proven successful in other contexts~\cite{shokri2017membership,tian2022comprehensive}. However, for speaker anonymization, we argue that using a model to identify relevant patterns in speech to de-anonymize speakers is not the best strategy. A better strategy is to design principled inference attacks and use those as lower bounds to quantify privacy. To demonstrate this, we propose a simple (ML-free) de-anonymization attack that leverages the specifics of the transformation of the original x-vector into the pseudo x-vector. We show empirically that this attack significantly outperforms current ML-based approaches based on training speaker verification/identification models. 

Our proposed attack is not only simpler and more accurate than existing alternatives, it is also more computationally efficient as it does not require training any machine learning models. Moreover, our attack is able to detect if the target speaker is not within the set of considered suspects, so it still infers information in such cases. Our results call for re-aligning evaluation of the privacy-utility tradeoff for x-vector speaker anonymization. Machine learning is a powerful, but it should {\em not} be used as the sole strategy for analyzing privacy-utility tradeoffs.

%% file: background.tex
\section{Background \& Related Work}
\subsection{VoicePrivacy Challenge}
As voice-based technologies become pervasive~\cite{chung2018voxceleb2,khalil2019speech}, 
concerns about protecting personal information embedded in speech \cite{nautsch2019preserving} 
have become increasingly urgent. In response, the VoicePrivacy Challenge~\cite{tomashenko2022voiceprivacy,tomashenko2024voiceprivacy,panariello2024voiceprivacy} offers a platform for researchers to explore and compare state-of-the-art methods to protect a speaker’s identity while preserving critical linguistic content.\footnote{\url{https://www.voiceprivacychallenge.org/}} In the VoicePrivacy Challenge, anonymization techniques are evaluated from both privacy and utility perspectives. Typically privacy performance is evaluated using the Equal Error Rate (EER) of a speaker verification/recognition model, which indicates how successfully a method prevents speaker re-identification, whereas utility is evaluated using Word Error Rate (WER) and Unweighted Average Recall (UAR) comparing the anonymized speech transcript to the original speech transcript. The VoicePrivacy Challenge also provides several baseline systems to guide researchers, the most popular of which is the class of x-vector speaker anonymization approaches~\cite{srivastava2020design,srivastava2022privacy,champion2021study}.

\subsection{X-Vector Speaker Anonymization}
A x-vector based speaker anonymization system~\cite{tomashenko2022voiceprivacy,tomashenko2024voiceprivacy} first extracts x-vectors, fundamental frequencies, and bottleneck features from original speech. It then anonymizes the speaker's identity by replacing the original x-vector with an average of numerous vectors selected from the public x-vector pool, creating a {\em pseudo-speaker x-vector}. Finally, using a neural source-filter model, it synthesizes a new speech waveform that retains the original's linguistic content but sounds as though it was uttered by a different individual.

Fang et al.~\cite{fang2019speaker} introduced the first x-vector based speaker anonymization method through voice conversion, which adapts the x-vector of a speaker to match a target x-vector. Following this, a number of other strategies were proposed~\cite{yoo2020speaker},~\cite{mawalim2020x},~\cite{meyer2023anonymizing} that include random modification of the embeddings, Singular Value Decomposition, and Wasserstein GAN to generate the target x-vector. 

Champion et al.~\cite{champion2021invertibility} performed linkability and invertibility attacks on anonymized x-vectors produced using the baseline system of the VoicePrivacy challenge \cite{tomashenko2020introducing}. This work used two different embedding alignment algorithms to evaluate x-vector based anonymization in scenarios where the attacker was completely informed or semi-informed about the original x-vector and its corresponding anonymized x-vector. Unlike ours, they use a machine learning based method. Champion et al.~\cite{champion2021evaluating} also analyzed x-vector based speaker anonymization proposed by~\cite{fang2019speaker} where the attacker has complete knowledge of the system.  

\begin{figure}
    \centering
    \includegraphics[width=1\linewidth]{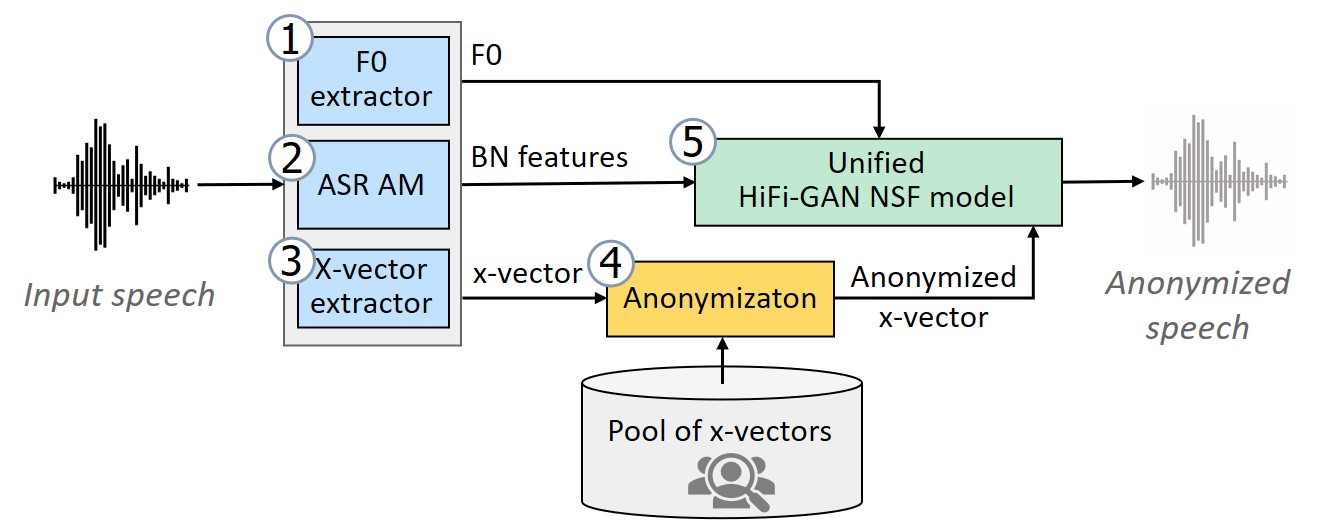}
    \caption{Model architecture for x-vector speaker anonymization.\protect \footnotemark}
    \label{fig:x-vector-model}
\end{figure}

There are different x-vector based speaker anonymization techniques. In this paper, we use baseline B1 in the 2024 VoicePrivacy Challenge (which is baseline B1.b in the 2022 challenge). \cref{fig:x-vector-model} shows its architecture. 
We consider it as our main representative x-vector anonymization technique. 
\footnotetext{Image source: {\url{https://github.com/Voice-Privacy-Challenge/Voice-Privacy-Challenge-2022/blob/master/baseline/fig/B1b.jpg} (License: Creative Commons Attribution 4.0 International)}}

\subsection{Privacy Evaluation}
Privacy evaluation in the VoicePrivacy Challenge~\cite{tomashenko2022voiceprivacy,tomashenko2024voiceprivacy,panariello2024voiceprivacy} assumes the attacker trains an automatic speaker verification (ASV) model on anonymized data. For each speaker, the attacker computes an average embedding from all anonymized enrollment utterances and compares it to the embedding of an anonymized trial utterance to verify speaker identity. They use EER as evaluation metrics. Srivastava et al.~\cite{srivastava2022privacy} evaluates privacy by assessing how well the speaker anonymization methods prevent an attacker from re-identifying the speaker using automatic speech recognition (ASR) model. The privacy evaluation is quantitatively measured using linkability scores, which reflect the likelihood of correctly linking anonymized speech to the original speaker. It outlines various attacker knowledge levels, ranging from ignorant (unaware of anonymization) to semi-informed (aware of some details of the anonymization method but not others), influencing how the attacker might use the anonymized data to re-identify the speaker. Champion et al.~\cite{champion2021evaluating} also evaluate the privacy of x-vector based speaker anonymization but in a white-box setting when the target selection is restricted to a specific identity. The privacy evaluation is performed using the linkability metric with an x-vector-PLDA based Automatic
Speaker Verification (ASV) system from the VoicePrivacy Challenge.

In this paper, we build on the concepts presented in~\cite{srivastava2020design,tomashenko2024voiceprivacy} but propose a novel privacy attack by examining a more knowledgeable adversary than has been considered in related work, distinct from the framework used in \cite{srivastava2022privacy}. 

Furthermore, our main insight is that x-vector speaker anonymization can be attacked directly by leveraging how they construct the pseudo x-vector from the original speech. It is not necessary to train a complex ASI/ASV model and hope it learns to de-anonymize.

%% file: problem.tex
\section{Problem Statement}
We frame the problem of speaker anonymization as follows. The input is a speech sample $a$ from a speaker $s$ whose identity we aim to protect. We use a speaker anonymization method that produces an anonymized speech sample $y$ such that $y$ approximately preserves the transcript of $a$ but obscures speaker identity. Some features, such as gender, may still be identifiable, but the speaker should be indistinguishable from other speakers that share those features.\footnote{Some methods preserve gender by restricting the public pool of x-vectors used to construct pseudo-speaker x-vector to the same gender as the original speaker.} The adversary observes the anonymized speech sample $y$ and attempts to determine the identity of the speaker, i.e., de-anonymize (re-identify) them. 

\subsection{Formalizing Embedding-based Anonymization}
We think of embeddings being x-vectors but the formalization extends to other (potentially future) representations of an individual's voice features.
An Embedding-based Anonymization Scheme is a tuple $(\ext, \transf, \synth)$ of algorithms where:
\begin{itemize}
    \item  $\ext(a) \to (x, t)$: Extract takes as input a speech sample $a$ and outputs a text transcript $t$ (a natural language string) and a speaker embedding $x \in \mathbb{R}^k$ (the x-vector) where $k$ is the embedding dimension (e.g., $k=512$).
    \item $\transf(x) \to p$: Transform takes as input a speaker embedding $x$ and transforms it (anonymizes it) into a different pseudo speaker embedding $p$. 
    \item $\synth(t, p) \to y$: Synthesize takes as input a text transcript $t$ and a pseudo speaker embedding $p$ and produces an audio speech sample $y$ as output.
\end{itemize}

This captures the idea and anonymization process of existing x-vector speaker anonymization schemes such as~\cite{tomashenko2020voiceprivacy,tomashenko2022voiceprivacy,tomashenko2024voiceprivacy}. That is, given an audio sample $a$ to anonymize, we use the function $\ext(a)$ to extract its transcript $t$ and x-vector $x$, then transform this x-vector into a {\em pseudo} x-vector $p$ using $\transf(x)$, and then finally use the function $\synth(t, p)$ to synthesize a new audio sample that matches the characteristics of the speaker represented by $p$. 

There are a few important remarks. Since x-vector transformations rely on a public pool of embeddings, we can think of $\transf$ as including this pool implicitly. To capture any randomness in the process of transforming an x-vector into another {\em pseudo} x-vector, we can think of $\transf$ as having an auxiliary input $r$ which is a source of randomness. For the purposes of thinking about formal security, we can even assume that $r$ is derived (for each invocation) from a cryptographic secret key, which is equivalent to thinking of the randomness in $\transf$ as coming from a PRNG seeded with the secret key. To simplify the presentation we omit this from the description and view $\transf$ as probabilistic.

In our representative scheme, the VoicePrivacy Challenge's x-vector anonymization implementation, the extract function actually produces a tuple $(x, F_0, B_N)$ where $x$ is the x-vector (embedding), $F_0$ is the pitch of the speaker, and $B_N$ represents the features of the transcript. Further, the synthesis function $\synth$ takes as input an pseudo x-vector $p$ in addition to $F_0$ and $B_N$, and therefore it is implicitly assumed that {\em no information about speaker identity} is contained in $F_0$ and $B_N$. 
Consistent with related work Shamsabadi et al.~\cite{shamsabadi2022differentially} we found that this is assumption is false empirically. We discuss this in~\cref{sec:res:idios}.

\subsection{Privacy}
A natural way to perform a de-anonymization attack is to try to (approximately) invert the transformation $\transf(\cdot)$. If we observe an anonymized speech sample with some pseudo-speaker embedding $p$, we want to identify the most likely speaker embedding $x$ such that $\transf(x) \to p$. This can be viewed as an inference attack where the attacker guesses based on some likelihood ratio or approximates it. For example, suppose we have x-vectors $x_1$ and $x_2$ from two distinct possible speakers and an (estimated) pseudo x-vector $p$ obtained from the anonymized audio. The attack computes ratio: $\pr{{\transf(x_1)} = p} / \pr{{\transf(x_2)} = p}$. 
The attacker guesses the first speaker if and only if the ratio is larger than $1$. The challenge is therefore to compute or approximate the ratio. This is the approach of the attack we propose in~\cref{sec:method}, although we avoid explicitly computing probabilities and use x-vector distances instead. 

By contrast, the approach taken in the VoicePrivacy challenges~\cite{tomashenko2020voiceprivacy,tomashenko2022voiceprivacy,tomashenko2024voiceprivacy} and related work is to perform de-anonymization by training an Automatic Speaker Verification (ASV) model. This model is then fed the anonymized speech sample and asked to predict the identity of the speaker. This is how various methods for the challenge are evaluated and compared to each other in terms of privacy. There are a number of downsides with this ML-based approach. In particular, it does not utilize fine-grained information about the anonymization method ($\transf$) or its invertibility. It requires training the ASV model which is computationally intensive and the performance of the attack depends on how well that model can learn speaker identity from (anonymized) speech samples.
\begin{figure*}[t]
     \centering
     \includegraphics[width=1\linewidth]{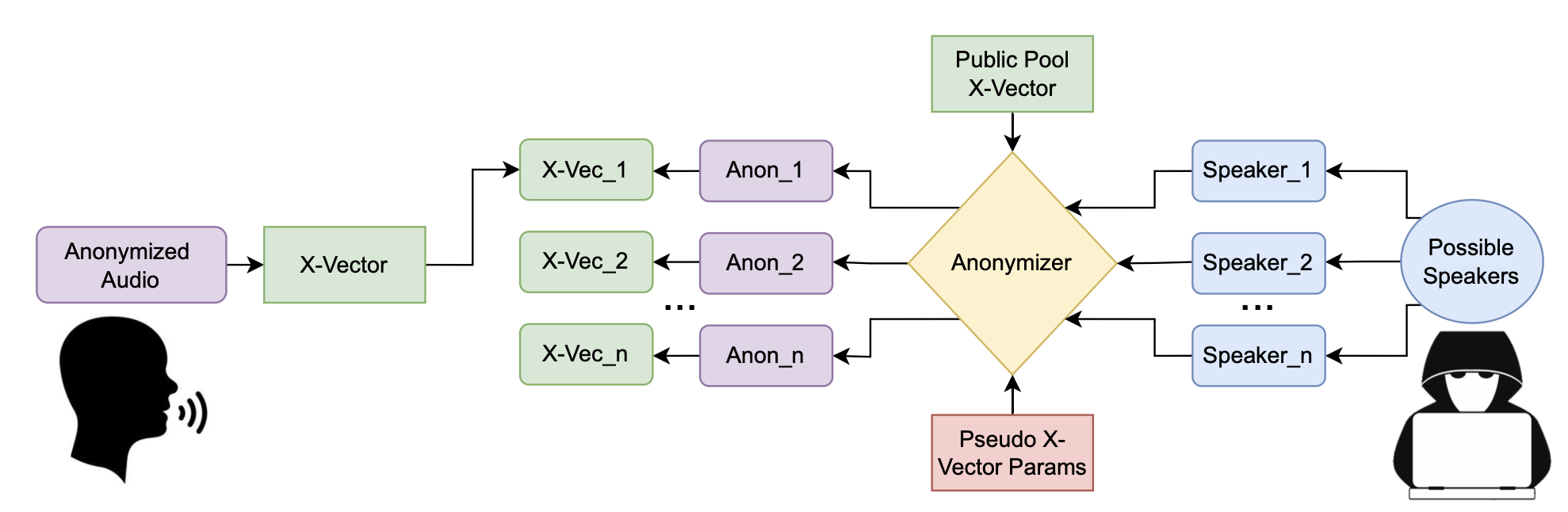}
     \caption{Illustration of our proposed attack. The target speaker, $s$, has released anonymized audio that the adversary is attempting to identify. The adversary has a set of potential speakers, $S'$ and they believe the speaker is a part of it. They anonymize audio from these potential speakers. Finally, they extract an x-vectors from the target anonymized audio $x$ and from the set of anonymized audio they just generated $X'$. The speaker in $S'$ whose x-vector most closely matches $x$ is the speaker of the target audio.  }
     \label{fig:attack}
 \end{figure*}
\subsection{Utility}
Privacy is not the only important criterion for speaker anonymization. Otherwise we could simply transcribe the original speech sample and then synthesize a new audio sample for that transcript with a fixed voice (e.g., robotic voice) independent of the original speaker. However, this would not preserve features such as gender, age~\cite{prajapati2021voice}, emotion~\cite{cai2024privacy}, or tone/speaking style, etc.

In the literature and the VoicePrivacy challenge, utility is often measured as distortion to the transcript (differences in transcription from the original to the anonymized audio), which are measured using WER (Word Error Rate). We also measure FAD (Fr\'echet Audio Distance)~\cite{kilgour2018fr}, which compares generated audio against a ground truth set of non-generated audio to determine quality. 

%% file: methodology.tex
\section{Methodology}\label{sec:method}

\subsection{VoicePrivacy Challenge}
Consider how the VoicePrivacy Challenge anonymizes audio~\cite{tomashenko2020voiceprivacy,tomashenko2024voiceprivacy}. The system first extracts an x-vector, and other data, such as pitch ($F_0$) and linguistic features ($B_N$), from the original audio. It then creates a \emph{pseudo x-vector} which is given, along with the $F_0$ and $B_N$ collected from the original audio, to a Neural Source Filter model. The model uses these to create the final anonymized audio. 

The system uses a public pool of x-vectors, which we denote as ${\rm pool}$. This pool is constructed from a large number of x-vectors gathered from independent audio. The VoicePrivacy Challenge uses the ``train-other-500'' subset of LibriTTS~\cite{zen2019libritts} as the source of this pool. There are thousands of utterances from about 500 speakers. Finally, the pool is divided by gender, which provides the ability to use either the same or opposite gender pool as the original speaker. In our experiments, we assume same gender pools. 

To construct a pseudo x-vector, the affinity between the original audio's x-vector and every other vector in the gender filtered pool is calculated. Affinity is a distance between vectors, and can be calculated either using the cosine distance or using PLDA. The list of affinity values is then sorted and then the top 200 vectors are selected. This may be the top 200 nearest vectors (highest affinity) or the 200 farthest vectors (lowest affinity) from the original x-vector. From there a subset of 100 vectors are randomly selected, and then averaged together. This average vector is the {\em pseudo x-vector}. This pseudo x-vector is then applied at either the speaker level (all utterances of a speaker get the same x-vector) or the utterance (each utterance gets its own x-vector). We consider speaker-level for our experiments. Finally, the pseudo x-vector is provided as input to the generation model, along with $F_0$ and $B_N$, to synthesize the anonymized audio.
\begin{algorithm}[!t]
  \caption{De-Anonymization Attack} 
  \label{alg:reconstruct}
  \small
  \begin{algorithmic}[1]
    \Require
      $ {y}$: Anonymized audio files from target speaker $s$;

      $ {\rm pool}$: Public pool of x-vectors;

      $  S'$: Set of potential speakers;

      $ A'$: Original utterances from speakers in $S'$

    \Ensure
    $s_i'$ in $S'$ that is most likely to be $s$

    \Procedure{Extract\_xvector}{$\rm audio$}
        \State Given several samples of $\rm audio$ from a speaker extract an x-vector for that speaker. 
    \EndProcedure

    \Procedure{Anonymize}{${\rm audio}, {\rm pool}$}
        \State Run representative x-vector anonymization method to obtain anonymized audio.  
    \EndProcedure

    \State $x \gets$ Extract\_xvector($y$)
    
    \For {$s_i'$ in $S'$}
        \State $y_i' \gets$ Anonymize(${a_i'}, {\rm pool}$)
        \State $x_i'$ $\gets$ Extract\_xvector($y_i'$)
        \State ${\rm dist}_i \gets ||x_i' - x||_2$


    \EndFor

    \State\Return{$\rm s'_i$ with the lowest dist}

  \end{algorithmic}
\end{algorithm}
\subsection{Proposed Inference Attack}
To reiterate, a speaker $s$ has several audio files $a$ they wish to anonymize. For example, these samples may consist of the speaker reading several sentences from a book, each sentence being a separate audio file also described as an ``utterance''. The speaker anonymizes $a$ to get new anonymized audio $y$. The adversary observes the anonymized speech sample $y$ and attempts to guess the identity of the speaker. The adversary has access to the anonymization method, as well as a set of potential speakers $S'$, for which they have audio samples $A'$. They believe that $s$ is within $S'$, and they attempt to identify them. 

Our proposed inference attack works as follows. First, the attacker extracts an x-vector $x$ from the observed anonymized speech sample $y$. This is done using the x-vector extractor from the VoicePrivacy Challenge, based on Snyder et al.~\cite{snyder2017deep}. In principle, this extracted x-vector should be similar, if not identical, to the pseudo x-vector used to create it. In practice, there are significant differences due to the audio generation and x-vector retrieval process. We will discuss this problem later. The attacker then simulates the anonymization process for each speaker in $S'$. From this, the attacker obtains anonymized audio for each speaker in $S'$ from which they extract x-vectors. This yields a set $X'$ of x-vectors where each x-vector $x_i' \in X'$ represents the speaker embedding/identity of $s_i'$. Finally, the attacker compares each $x_i'$ to $x$ to find the speaker most similar to the original target $s$ (using $l_2$ distance). \cref{alg:reconstruct} shows details of the attack method.

This attack works by leveraging the pseudo x-vector construction method. One may expect the anonymized audio to contain no information about the identity of the original speaker, since it is constructed from the pseudo x-vector. However, the pseudo x-vector is not constructed independently from the original speaker x-vector, thus it carries statistical information from it. The specific steps of selecting the 200 nearest/farthest x-vectors (to the original speaker x-vector) from the pool leaks information about it through the affinity/distance. In the ideal case for the adversary, the 200 nearest/farthest x-vectors acts as fingerprint for the speaker identity $s$. In such cases, the inference attacks only needs to overcome the uncertainty of the 100 randomly selected x-vectors from within the 200 nearest/farthest x-vector set.

\begin{table*}[!t]
\caption{Attack accuracy and utility for different pseudo x-vector construction methods. We show accuracy for our main attack method under the Same and Different adversary models, as well as an ASI model using Same, Different, and Original adversary models. For utility, we show the WER and FAD of all pseudo x-vector methods. Random Single has the lowest attack accuracy and best utility.}
\centering
\resizebox{0.8\textwidth}{!}{%
\begin{tabular}{c|cc|ccc|cc}
\toprule
 & (Ours) Same & (Ours) Different & ASI Same & ASI Diff & ASI Orig & WER & FAD \\ \midrule
Original Audio & 100 & 96.7 & 79.3 & 70.1 & 79.3 & 5.4 & 2.1 \\
200 Farthest & 100 & 76.3 & 34.5 & 19.6 & 9.2 & 7.4 & 7.5 \\
200 Nearest & 100 & 77.6 & 43.6 & 20.2 & 12.6 & 7.8 & 7.2 \\
50 Farthest & 100 & 65.7 & 32.4 & 18.2 & 11.1 & 7.7 & 7.5 \\
50 Nearest & 100 & 73.9 & 31.5 & 17.8 & 14.7 & 7.6 & 6.9 \\
Random Average & 100 & 46.3 & 36.3 & 18 & 9.8 & 6.4 & 7.4 \\
Random Single & 20 & 11.2 & 33.2 & 14.6 & 5.3 & 6.9 & 6.5 \\ \bottomrule
\end{tabular}
}
\label{tbl:attack}
\end{table*}

%% file: results.tex
\section{Experiments}
\label{sec:results}

\subsection{Setup}
\begin{table}[tbh]
\caption{Pseudo x-vector construction methods. All consist of selecting a \emph{World} of nearest or farthest x-vectors from ${\rm pool}$, then averaging a random subset of them together. Random Single is the exception since it only uses a single x-vector instead of an average.}
\centering
\resizebox{0.905\columnwidth}{!}{%
\begin{tabular}{l|l}
\toprule
Method & Description \\ \midrule
\multicolumn{1}{l|}{200 Farthest} & Average 100 vectors out of 200 farthest \\
\multicolumn{1}{l|}{200 Nearest} & Average 100 vectors out of 200 nearest \\
\multicolumn{1}{l|}{50 Farthest} & Average 25 vectors out of 50 farthest \\
\multicolumn{1}{l|}{50 Nearest} & Average 25 vectors out of 50 nearest \\
\multicolumn{1}{l|}{Random Average} & Average 100 vectors of the entire ${\rm pool}$ \\
\multicolumn{1}{l|}{Random Single} & A single vector from the entirety of ${\rm pool}$ \\
\bottomrule
\end{tabular}
}

\end{table}
 
To evaluate our method we apply the VoicePrivacy Challenge anonymization process to the Libri\_Dev dataset~\cite{panayotov2015librispeech}. \texttt{Libri\_dev\_trials\_m/f} are considered the original pool of audio $A$ belonging to speakers $S$, which is then anonymized. \texttt{Libri\_dev\_enrolls} is used as the adversary's target pool, $S'$ when a separate and distinct pool is required. There are 29 unique speakers in $S'$, which is the number of unique speakers in \texttt{Libri\_dev\_enrolls}. For some experiments we use a smaller subset to show how the size of $S'$ influences attack accuracy. Each speaker $s_i$ is associated with $a_i$, composed of several utterances of the speaker reading sentences from a book. The exact number of utterances depends on the speaker.

We wish to examine how the pseudo x-vector generation parameters affect privacy and utility. We discussed how the VoicePrivacy Challenge constructs pseudo x-vectors in Section \ref{sec:method}, by gathering a \emph{World} of 200 nearest / farthest x-vectors from which 100 are randomly selected. We consider both settings to evaluate our representative method. We further evaluate a smaller \emph{World} of 50 from which 25 are sampled, to show how the size of the pool potentially influences the privacy of the speakers.
Finally, we also evaluate two additional scenarios that should provide maximum privacy. The first takes the average of 100 randomly selected x-vectors from {\em the entire pool} and uses it as the pseudo x-vector. The second selects a single x-vector from the pool randomly and uses it as the pseudo x-vector. Since the selection of pseudo x-vector in those two scenarios is not dependent on the original speaker or speech, there should be no leakage.

\subsection{Attack Scenarios}
 \begin{table}[tbh]
\caption{Different adversary knowledge levels we evaluate. We evaluate Same, Different, and Unknown for our attack. We evaluate Same, Different, and Original for the ASI model.}
\label{table:knowledge}
\centering
\resizebox{1\columnwidth}{!}{%
\begin{tabular}{l|l}
\toprule
Knowledge & Description \\ \midrule
\multicolumn{1}{l|}{Same} & The target's original non-anonymized audio is in $A'$ \\
\multicolumn{1}{l|}{Different} & The target speaker $s$ is in $S'$ but with different utterances \\

\multicolumn{1}{l|}{Original} & (ASI only) Original non-anonymized audio to train model \\
\bottomrule
\end{tabular}
}
\end{table}
Recall that in our threat model the adversary knows the anonymization method, the pseudo x-vector construction method, and any other parameters. As a result, the adversary can replicate steps taken during the anonymization process, such as x-vector extraction and pseudo x-vector construction. The adversary also has access to a set of potential speakers $S'$, for whom the adversary has original non-anonymized audio samples $A'$. 

In~\cref{table:knowledge} we propose three attack scenarios that map onto different adversarial knowledge levels, representing how closely the adversary's audio samples mirror the samples used by the anonymization method. Our most powerful attack assumes that the adversary knows the original speaker is within that set and that the audio samples are the same as the ones used to generate the target audio $y$, i.e. $A'=A$. We also evaluate a weaker, but perhaps more realistic scenario, where $S'$ contains the target speaker, but with different audio samples, i,e, $A' \neq A $. For the ASI model, we also evaluate the scenario where the adversary does not have any anonymized audio and trains only on the original audio. We further evaluate the scenario where the adversary does not know if the target speaker is within $S'$.

\paragraphb{Attacks}
We evaluated our proposed inference attack, implemented as described in~\cref{sec:method}. As a comparison point, we use an Automatic Speaker Identification (ASI)-based attack, which attempts to identify the original speaker by examining anonymized audio samples rather than the x-vectors. More specifically, we train an ASI neural network~\cite{Ruwali_2020} using data corresponding to the assumed adversary knowledge levels and perform inference to re-identify the target speakers. We train until loss stops decreasing, usually around 10 epochs. We use the \texttt{Libri\_dev} male and female subsets as training samples. Related work often relies on an ASV (which takes in audio files and a claimed identity to verify the speaker) for this. Instead, we opted to use a speaker identification model since it better fits our attack scenario, having a set of audio files and attempting to classify each one as a certain speaker.

\subsection{Results}\label{sec:exp:res}
\cref{tbl:attack} gives the results for our attack at different adversary knowledge levels and pseudo x-vector construction methods. We find that at the Same adversary knowledge level, we are able to achieve perfect de-anonymization accuracy at all sizes of $S'$ and for (nearly) all generation parameters. Random Average and Random Single achieve better accuracy than expected, due to leakage of $F_0$ and $B_N$ features in the x-vector extraction process, which we further discuss later on. 

\begin{figure}[t]
     \centering
     \includegraphics[width=0.8\linewidth]{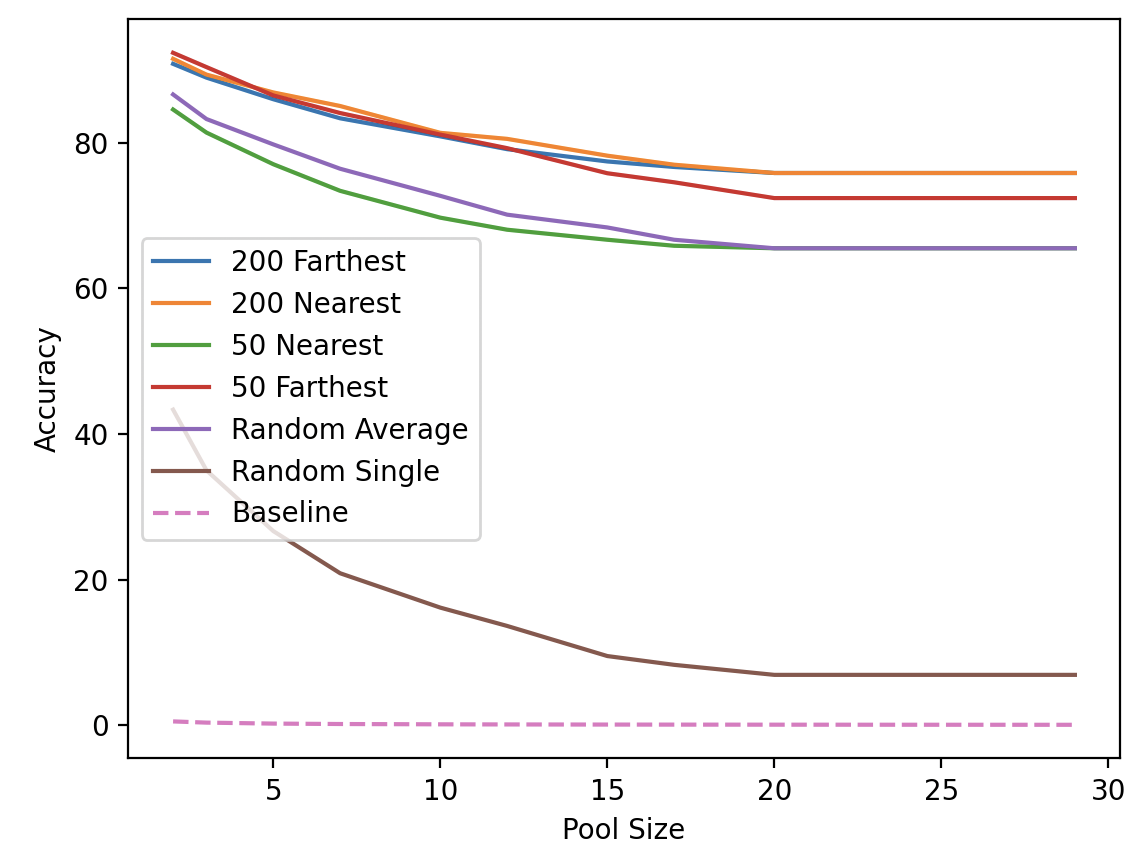}
     \caption{Attack accuracy for different pseudo x-vector construction methods under the Different adversary knowledge level. All construction methods, except for random, perform much better than random guessing. As the adversary's pool size increases, the attack accuracy decreases before leveling out around a pool of size 20.
     }
     \label{fig:different}
\end{figure}

Results for the Different adversary knowledge scenario are shown in~\cref{fig:different}. Arguably this is a more realistic adversary scenario, and is reflective of the most knowledgeable adversary considered in related work. Nevertheless, our inference attack still achieves high re-identification accuracy. 

\cref{tbl:attack} also shows the results for the ASI at various adversarial knowledge levels. As expected the model performs better with stronger adversaries. Perhaps, due to the small sample sizes for each speaker in \texttt{Libri\_dev}, it is unable to achieve high accuracy. 

Overall results suggest that training complex deep neural networks as an attack is unnecessary, as our (ML-free) inference attacks performs as well or better without training any model. It only requires knowledge of the anonymization method and the public pool of x-vectors. Furthermore, given the high success rate, the anonymization methods considered do not appear to provide meaningful privacy.

\subsection{Open World Evaluation}
We also evaluated the open world setting where it is not known if the original speaker $s$ is within the set $S'$ of possible targets. The question in this case is whether the attack is resilient to the possibility that the original speaker is not in the target set.

To evaluate this, we use our inference attack, but we only include the target speaker $s$ within $S'$ with probability $0.5$ each time. Our attack accounts for this as follows. Instead of sorting possible speaker by their distance to the x-vector uses to construct the anonymized audio, the attack compares the minimum distance against a threshold. If the distance is smaller than the threshold then the adversary will declare the target is within the set. To determine a suitable threshold, the adversary can perform the attack in a setting where the speaker $s$ is within the set $S'$ to estimate the distribution of x-vector $l_2$ distances and then derive a threshold value from it (e.g., choosing a threshold at or above a given percentile). 

\begin{figure}[t]
     \centering
     \includegraphics[width=0.9\linewidth]{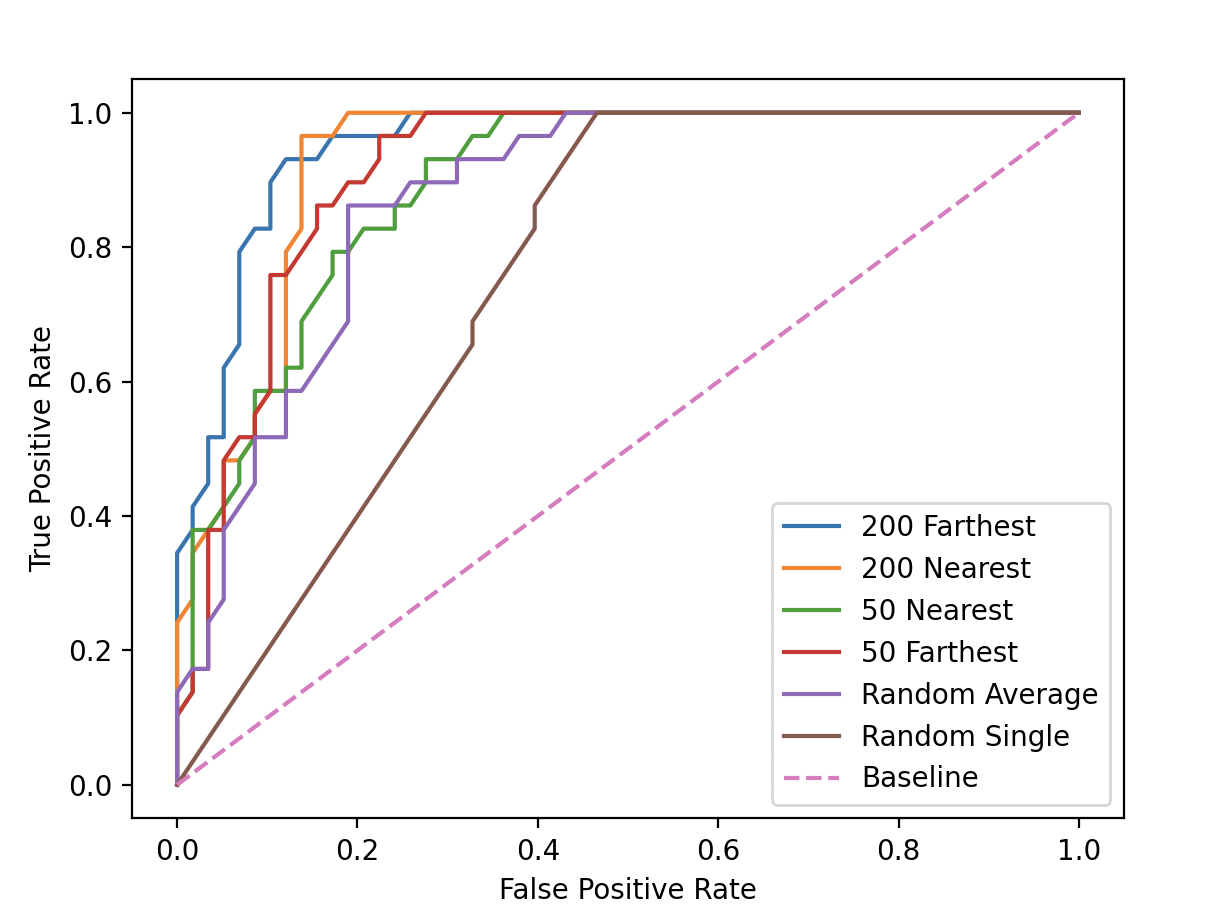}
     \caption{ROC curves for identifying if a speaker is within the potential target pool when anonymized using different pseudo x-vector construction methods. All results shown are under the Different adversary knowledge level. AUC is far above random guessing for all. Not shown is the Same adversary knowledge level, which has AUC=1 for all pseudo x-vector construction methods, except Random Single, which is still above random guessing.}
     \label{fig:roc}
\end{figure}

We found that in the Same adversary knowledge setting we again get perfect accuracy, i.e., we are able to select a threshold without any false positives or false negatives. \cref{fig:roc} shows the ROC curves for the Different adversary knowledge scenario. We observe that even in this case, the adversary is able to identify if the speaker is present with high accuracy. 

\subsection{Other Experimental Results}
\paragraphb{Time Comparison}
Anonymizing \texttt{Libri\_dev} took approximately 21 minutes, most of it to extract all the necessary information and generate the anonymized audio. Note that both attacks methods (ours and ASI/ASV) must perform this step. However, from there it took approximately 12 minutes to train the ASI model to convergence, solely on \texttt{Libri\_dev}. Note that the ASV used in the VoicePrivacy Challenge can take up to 10 hours to train on their recommended dataset. However, the only time consuming step in our attack is extracting x-vectors from \texttt{Libri\_dev}, which only takes approximately 2 minutes. Our attack is thus much faster, and needs no training or additional models. 

\paragraphb{Utility}
To evaluate how the different generation parameters influenced utility, we compare anonymized audio against non-anonymized audio using WER and FAD. While utility is maintained across methods, Random Single has high utility across both metrics. Our results suggest Random Single is the best option, since it also achieves the highest privacy across all knowledge levels.

\begin{table}[th]
\caption{Results for our normalized scenario, where the generator is given the same F0 and BN regardless of the input audio. Hence the only difference between audio files is the x-vector used to generate it. We also show the accuracy for our Same and Different knowledge level attacks for comparison.}
\centering
\label{tbl:normalized}
\resizebox{0.9\columnwidth}{!}{%
\begin{tabular}{c|ccc}
\toprule
 & Same & Different & Normalized \\ \midrule
\multicolumn{1}{l|}{Original Audio} & 100 & 96.7 & 100 \\
\multicolumn{1}{l|}{200 Farthest} & 100 & 76.3 & 93 \\
\multicolumn{1}{l|}{200 Nearest} & 100 & 77.6 & 93.1 \\
\multicolumn{1}{l|}{50 Farthest} & 100 & 65.7 & 79.3 \\
\multicolumn{1}{l|}{50 Nearest} & 100 & 73.9 & 99.6 \\
\multicolumn{1}{l|}{Random Average} & 100 & 46.3 & 3.5 \\
\multicolumn{1}{l|}{Random Single} & 20 & 11.2 & 3.4 \\ \bottomrule

\end{tabular}
}
\end{table}

\subsection{Idiosyncrasies of X-Vector Anonymization}\label{sec:res:idios}

Recall that Random Single and Random Average should have no privacy leakage. Yet, the results (\cref{tbl:attack}) show that our attack on these methods achieve well above random guessing accuracy. This should not happen because the pseudo x-vectors should contain no information about the original speaker. We discovered that the reason for the empirical outperformance is leakage of information about $F_0$ and $B_N$ into the anonymized audio. We believe this was also observed indirectly by Shamsabadi et al.~\cite{shamsabadi2022differentially}.

To evaluate how this affects our attack success, we performed the attack again while forcibly setting the $F_0$ and $B_N$ features to ensure no leakage from them. This means both $y$ and $y'$ will have the same utterances, pitch values, and parameters, except for the pseudo x-vector used to generate them. We call this the {\em normalized} scenario. Results are shown in~\cref{tbl:normalized} where we see that (as expected) both Random Average and Random Single achieve only random guessing accuracy. Recall the size of $S'$ is based on \texttt{Libri\_dev\_enrolls}, which contains 29 unique speakers. Therefore, random guessing accuracy is $3.45\%$. Our attack on other methods also sees small decreases to accuracy, but can still easily identify the speaker most of the time.

%% file: limitations.tex
\section{Discussions \& Limitations}

\paragraphb{Discussion}
Although not the focus of our paper, we found empirically that is surprisingly easy to identify audio that has been anonymized with x-vector based methods when compared against non-anonymized audio. We fine-tuned a Whisper~\cite{radford2022robust} model to distinguish between audio anonymized with 200 farthest pseudo x-vectors and non-anonymized audio samples from the same dataset and found it easily reached perfect accuracy. This is likely due to a combination of averaged pseudo x-vectors resulting in very neutral sounding audio. We believe this is noteworthy, especially in light of the recent attention on deepfake audio~\cite{warren2024better,hamza2022deepfake}.
However, it is unlikely to be major issue for speaker anonymization except in scenarios where concealing that anonymization has taken place is essential.


Our results suggest that any x-vector anonymization that carries information from the original x-vector to the pseudo x-vector can be broken. However, preserving features such as tone, pitch, gender, etc. may be essential for utility. This is why approaches that achieve stronger privacy such as those based on differential privacy~\cite{shamsabadi2022differentially,yao2023dp}
or methods such as Random Single and Random Average may not be practical in many scenarios. Random Single, in particular, also has the potential drawback that it selects the x-vector of an actual speaker from the pool and thereby essentially results in impersonating that individual.


\paragraphb{Limitations}
Our machine learning experiments using speaker identification (ASI) to classify audio as belonging to one of several speaker. This makes sense in our setting. Nevertheless it would be appropriate in future work to consider the case of speaker verification (ASV) for the attack model, where a model is trained to recognize audio as belonging to a single speaker.



Another limitation is scalability, which we have not evaluated, especially in cases where the pool contains a very large number (e.g., millions) of distinct speakers. This is particularly challenging since our attack needs to compare the distances of all potential targets against the pool, which quickly becomes prohibitive as the pool size grows. Future work may consider optimizations to narrow down the potential pool, or speed up comparisons.


%% file: conclusions.tex
%
\section{Conclusions \& Future Directions}
We proposed a novel attack on x-vector speaker anonymization that does not require any additional model training. The existence of this attack and its outperformance of existing automated speaker identification attack approaches underlines the importance of considering the specifics of the anonymization process, and not just the end result audio hoping that attack models will learn to identify relevant patterns during training. 

Future research can build on our results by taking into account the invertibility of pseudo x-vector generation methods to optimize the privacy-utility tradeoff. Additional assessments of utility through user study experiments could further improve pseudo x-vector construction methods.